\documentclass[12pt]{article}

\usepackage[english]{babel} 

\begin{document}
\title{The lattices with the continious vorticity as a model for FQHE.}
\author{ S.V.Iordanski \\
Landau Institute for Theoretical Physics, RAS, \\
Russia 142432 Chernogolovka, pr. Ac.Semenov, 1a
}
\maketitle
\begin{abstract}
It was shown that the including  spin of 2d electrons at high magnetic field is possible to remove the divergencies in the
cores of the vortex lattice and construct the topologically stable states. These states can be considered as the lattices of skyrmions where the unit cell is mapped on the whole sphere of spin directions.That gives the gapped ground state for
electrons and can be used as a model for FQHE.
\end {abstract}

The experimental discovery of the integer Quantum Hall Effect (IQHE) (K. von Klitzing (1980)) 
and the Fractional ( FQHE) 
(Tsui, Stoermer, Gossard (1982)) were among the most outstanding achievements in condensed matter physics of the last century. A qualitative theory for FQHE was developed in a short time after its discovery and is connected with the existence
of the energy gap at the full filling of Landau level. The theory of FQHE is more complicated because even a qualitative
description is inpossible in one electron picture (see e.g. reviews \cite{Prange,Das}). A large number of theoretical works
was based on the assumption that the ground state of 2d electrons at high magnetic field can be constructed from one electron states on the lowest Landau level  (e.g the famous Laughlin wave function \cite{Laugh} ). This assumption does not take into account the full thermodymic energy neglecting the possibility of the electrical currents in the ground state. The most successful phenomenological theory of "composite" fermions explains some part of the observed fractions for FQHE by the existence of the "additional" magnetic  flux carried by 2d electrons \cite{Jain}, but gives no possibility
to compute the energy gaps.

 In the recent works of  the author\cite{IordanP},\cite{IordanJP} was shown that the projected states are unstable to the vortex formation which can decrease the thermodymical energy irrespective to the details of the construction and the electron interaction. The vortex formation gives the current with some magnetic moment decreasing the thermodynamic energy.
 The arising vortex lattice imitate the "additional" magnetic flux anticipated by the Jains theory of "composite" fermions
 \cite{Jain}. However no specified model was considered in \cite{IordanP},\cite{IordanJP}. In this work I show that taking into account spin degrees of freedom for electrons gives the model of the vortices with a "soft" core a bit reminiscent of 
 the soft core vortices in the superfluid $^3He$ \cite{Salomaa}.

In a strong magnetic field the ground state of 2d electrons is ferromagnetic because of Stoner effect due to the state
degeneration , e.g. at the partial filling of the Ll. It is reasonable to suggest that this ferromagnicity is valid also on the local level for the nonuniform states. Therefore I can make a canonical thansformation of the second quantized fermi operators (spinors) $\Psi, \Psi^+$
$$\Psi =U\chi\: ,\:\Psi^+=\chi^+U^+$$
where $U=U_z(\gamma)U_y(\beta)U_z(\alpha)$ is the rotation matrix with 3 Euler angles depending on the electron position. Because of the unitarity of $U$, it is easy to show that if $\Psi_\sigma(r), \Psi _\sigma^+(r)$ have fermi commutations, then $\chi_{\sigma},\chi_{\sigma}^+$ also are fermions and vice versa. Any phsical operator can be transformed to a new representation. The interaction term in the hamiltonian depends only on the density operators and therefore is the invariant of this transformation.Indeed the transformations must be performed only for the electrical current operator and the operator of the kinetic energy. 

The current operator reads
\begin{eqnarray}
\label{current}
j=\frac{\hbar e}{2m}(-\imath\Psi^+ {\bf  \nabla}\Psi+\imath{\bf \nabla}\Psi^+\Psi-\frac{2e}{c\hbar}{\bf A})=\nonumber \\ 
=\frac{\hbar e}{2m}(-\imath\chi^+ {\bf  \nabla}\chi+\imath {\bf  \nabla}\chi^+\chi-\frac{2e}{c\hbar}{\bf A}+2\chi^+\Omega\chi) 
\end{eqnarray}
 where $\Omega =-\imath U^+ {\bf  \nabla} U$, ${\bf A}$ is the external vector-potential, giving the external magnetic field.

The standard expression for the rotation matrix \cite{LL3} is
\begin{equation}
U=\left(\begin{array}{cc}
\cos\frac{\beta}{2}e^{\imath\frac{\alpha+\gamma}{2}}&\sin\frac{\beta}{2}e^{-\imath\frac{\alpha-\gamma}{2}}\\
-\sin\frac{\beta}{2}e^{\imath\frac{\alpha-\gamma}{2}}&\cos\frac{\beta}{2}e^{-\imath\frac{\alpha+\gamma}{2}}\\
\end{array}
\right)
\end{equation}
It is easy to calulate the matrix $\Omega$
\begin{equation}
\label{Omega}
\Omega=\frac{1}{2}( {\bf  \nabla} \alpha+\cos \beta  {\bf  \nabla}\gamma)\sigma_z+\frac{1}{2}(-\imath {\bf  \nabla} \beta+ {\bf  \nabla} \gamma\sin \beta)e^{-\imath \alpha}\sigma_{+}+\frac{1}{2}(\imath {\bf  \nabla} \beta+ {\bf  \nabla}\gamma\sin \beta)e^{\imath \alpha}\sigma_{-}
\end{equation}
and

$$\sigma_{+}=\frac{1}{2}(\sigma_x + \imath\sigma_y)$$,
$$\sigma_{-}=\frac{1}{2}(\sigma_x-\imath\sigma_y)$$

where $\sigma_x,\sigma_y,\sigma_z$ are Pauli matrices.

In the same way we can transform the kinetic energy term
\begin{eqnarray}
\label{kinenergy}
H_T=\frac{\hbar^2}{2m}\int \chi^+\left((-\imath  {\bf  \nabla} -\frac{e}{c\hbar}{\bf A}+\Omega_z\sigma_z)^2+\frac{( {\bf  \nabla}\beta)^2}{4}+\sin^2\beta\frac{( {\bf  \nabla}\gamma)^2}{4}\right )\chi d^2r+\nonumber \\
+\frac{\hbar^2}{2m}\int\chi^+\left((-\imath  {\bf  \nabla} -\frac{e}{c\hbar}{\bf A})\Omega^{nd}+\Omega^{nd}(-\imath  {\bf  \nabla} -\frac{e}{c\hbar}{\bf A})\right)\chi d^2r 
\end{eqnarray}
where $\Omega^{nd}$ denote the nodiagonal part of the eq (\ref{Omega}).

Thermodynamic energy of 2d electrons at the given external vector-potential ${\bf A}$ reads \cite{LL8,IordanJP}
\begin{equation}
\label{tenergy}
F=H_i-\frac{1}{c}\int{\bf Aj} d^2r
\end{equation}
where
\begin{equation}
H_i=H_{T}+\frac{1}{2}\int V_c({\bf r-r'})\Psi^+({\bf r})\Psi^+({\bf r'})\Psi({\bf r'})\Psi({\bf r})d^2rd^2r'
\end{equation}
Here we omitted the interaction with the positive charged donors, meaning only the exchange interaction.

The rotation marix $U$ must have nontrivial topolgical properties, correspoding to the assumption about the vortex lattice being also a regular and simple function.  Therefore we assume 
\begin{equation}
\gamma({\bf r})=\alpha({\bf r})+\gamma'({\bf r})
\end{equation}
where $ {\bf  \nabla} \alpha$ has a vortex like singularities on a periodic lattice, the function ${\bf \nabla \gamma'({\bf r}) }$ is regular and periodic.The angle $\beta({\bf r})$ is also regular and periodic and is equal to $\pi$ on the vortex lattice of $ {\bf  \nabla} \alpha$, and is equal to 0 on the boundary of the unit cell of the vortex lattice.

It is easy to check that $U$ and $\Omega$ are regular and simple fuctions on 2d plain with the eliminated divergencies. We shall neglect the nondiagonal contributions giving the change of the spin direction for $\chi$. We consider only one direction $\chi_+=\left(\begin{array}{c} \chi \\ 0 \end{array}\right)$ for which the vortices decrease the termodynmic energy. Therefore $curl{\bf \Omega_z}$ has only one vector component orthogonal to 2d plane, and the sign is reversed compare to the external magnetic field $B$(also normal to 2d plain, because only such vortices decrease the termodynamic energy).

The nondiagonal terms neglecting can be justified by the large increase of energy due to the sign change of $\Omega_z$ and the imposibility to get new diagonal hamiltonian by the continuous change of the initial. It can be shown also that the nondiagonal terms have no terms transporting the states of the initial diagonal hamiltonian into the states of a new diagonal hamiltonian. In quantum field theory \cite{Radja} it is postulated that in thermodynamic limit there is no transition matrix elements between any different topologicals sectors. Later I show that in thermodynamic limit there is no real transitions
due to the nondiagonal terms.

Thus we get the effective hamiltonian in the form

\begin{eqnarray}
\label{effham}
H_{eff}=\frac{\hbar^2}{2m}\int \chi^+\left(
	\{(-i {\bf  \nabla}-\frac{e}{c\hbar}{\bf A}+\frac{1+\cos \beta}{2} {\bf  \nabla} \alpha+\frac{\cos\beta}{2} {\bf  \nabla} \gamma'\}^2+ \right. \nonumber \\
\left. +\frac{( {\bf  \nabla}\beta)^2}{4}+(\sin\beta\frac{ {\bf  \nabla}\alpha+ {\bf  \nabla}\gamma'}{2}\}^2 \right) \chi d^2r- \nonumber \\
- \frac{1}{c}\int{\bf A j }d^2r+\frac{1}{2}\int V_c({\bf r}-{\bf r'})\chi^+({\bf r})\chi^+({\bf r'})\chi({\bf r'})\chi({\bf r})d^2rd^2r' 
\end{eqnarray}
where 
\begin{equation}
\label{effj}
{\bf j}=\frac{\hbar e}{2m}\left( -i \chi^+  {\bf  \nabla}\chi{+i} {\bf  \nabla}\chi^{+}\chi+2\chi^{+}(-\frac{e}{c\hbar}{\bf A}+\frac{1+\cos \beta}{2} {\bf  \nabla}\alpha+\cos\beta \frac{ {\bf  \nabla}\gamma'}{2}\}chi\right)
\end{equation}

It is convenient to choose the representation of $ {\bf  \nabla}\alpha=(Im\zeta;Re\zeta)$ by  Weierstrass $\zeta$ function
\cite{WW}
$$\zeta(z)=\frac{1}{z}+\sum'_{n,n'} (\frac{1}{z-T_{n n'}}+\frac{1}{T_{n n'}}+\frac{z}{T_{n n'}^2})$$
where $T_{n n'}=0$ is excluded in the summation, $z=x+i y$,$T_{n n'}=\tau n+\tau' n'$ are complex periods. 
Alternatively  we can take the sum of the $\zeta$-functions with the shifted singularities. In any case $\zeta(z+\tau)=\zeta(z)+\delta(\tau)$, where $\delta$ does not depend on $z$.

If we choose $ {\bf  \nabla} \gamma = {\bf  \nabla} \alpha $, $ {\bf  \nabla}\gamma'=0$, then after the translation ${\bf r}\rightarrow{\bf  r}+\vec{\tau}$ the change is
$$ {\bf  \nabla} \gamma(r+\tau)= {\bf  \nabla}\alpha({\bf r}+\vec{\tau})= {\bf  \nabla}\alpha({\bf r})+\vec{\delta}(\vec{\tau})$$
according to the properties of $\zeta$-function.

This change can be eliminated if we choose 
\begin{equation}
 {\bf  \nabla}\gamma'+\vec{\delta}(\vec{\tau})=0
\end{equation}
In the same way the change of $\frac{1}{2}(1+\cos \beta ){\bf  \nabla}\alpha+\cos\beta\frac{ {\bf  \nabla}\gamma'}{2}-\frac{e}{c\hbar}{\bf A}$ can be eliminated by the change of the overall phase of $\chi$
\begin{equation}
 {\bf  \nabla} \phi-\frac{e}{c\hbar}{\bf A}(\vec{\tau})+\frac{ {\bf  \nabla}{\gamma'}}{2}=0
\end{equation}
The gauge corresponds to the linear function ${\bf A(r)}$.

Therefore the effective hamiltonian is invariant under the translation combined with the additional rotation on the angle $1/2\gamma'$ and the proper change of the phase of $\chi$. That is isomorphic to the group of magnetic translations (see e.g. \cite{LL9},\cite{B}).

It is known that finite representations of the ray group of magnetic translations exist only for the rational values of the total flux through the unit cell of the vortex lattice:
\begin{equation}
Bs-K\Phi_0=\frac{l}{N}\Phi_0
\end{equation}
where $s$ is the area, $K$ is the number of unit vortices in the unit cell, $\Phi_0=\frac{2\pi\hbar |e|}{c}$ is the flux quantum, $l$ and $N$ are integer. The proper representations of the magnetic translation group are well known and are enumerated by quasimomentum and for the filled band correponds to one electron per unit cell of the vortex lattice with $NM_1\times NM_2$ unit cells. Therefore the filled bands with the gaps on the boundaries exist only at electron density
\begin{equation}
n_e=\frac{1}{s}=\frac{N}{l+NK}\frac{B}{\Phi_0}
\end{equation}

All observed fractions correspondes to the validity of this formulae for the lattices with one or two unit vortices in the unit cell.

Using the representation of magnetic translation group it is possible to show that indeed in the thermodynamic limit the 
nondiagonal terms $\Omega^{nd}$ give the vanishing small contribution to the total energy. The second order of the perturbation theory in the non diagonal terms gives the correction to the ground state energy
\begin{equation}
\label{delta}
\delta E=\sum_{E({\bf k}),E'({\bf k'})}\frac{<\chi^{+}_{up,E}H^{+}\chi_{down,E'}><\chi_{down,E'}^{+}H^{-}\chi_{up,E}>}
	{E_{\bf k}-E'_{\bf k'}}
\end{equation}
where $H^{+}$ and $H^{-}$ are non diagonal terms increasing and decreasing  the spin projection. The one particle
energy difference $E({\bf k})- E'({\bf k'})$ does not contain the sample size and is not essential to the estimate. The 
energies $E({\bf k})$ correspond to the states in the filled up band, energies $E'({\bf k}')$  corresponds to the empty down
band. The angle brackets denote the space integral of the matrix transition element between the up- down states and the
inverse between down-up states. The nondiagonal parts $H^{+}$ and $H^{-}$  given by eq(\ref{current}),eq(\ref{Omega}),eq(\ref{kinenergy}),
eq(\ref{tenergy}) contain the nondiagonal parts $\Omega^{+}=(-i/2{\bf \nabla}\beta+1/2sin\beta{\bf \nabla} \vec{\gamma})\exp{(-i\alpha)}$ and $\Omega^{-}=(i/2{\bf \nabla}\beta+1/2sin\beta{\bf \nabla}\gamma)\exp{(i\alpha)}$.
The one particle wave functions have the standard Bloch representation
$$\chi_{up}=\sum_{\bf k} \exp{(i{\bf kr})}\frac{u_{{\bf k},up}({\bf r}) c_{{\bf k},up}}{\sqrt N}$$
$$\chi_{down}=\sum_{\bf k'}\exp{(i{\bf k'r})}\frac{u_{{\bf k'},down}({\bf r})c_{{\bf k'}, down}}{\sqrt N}$$
Here the functions $u_{up}({\bf r})$ and $u_{down}({\bf r})$ are periodic and normalized inside the unit cell, $N$ is the number
of the unit cells in the lattice. To avoid the cumbersome formulae I consider only the case of integer flux through the 
unit cell area. The only factors noninvariant to magnetic translations  in eq (\ref{delta}) are $ \exp({\pm i\alpha})$  contained in  a $\Omega^{\pm}$. Due to the properties of $\zeta$ function these factors give 
the strong oscillations inside the unit cells at the large distances  from the central cell near the origin.

The value of ${\bf \nabla}\alpha$ can be estimated using  Stokes theorem for a grand circle of radius $r=L\gg\sqrt s$ far
from the origin
$$\oint{\bf \nabla}\alpha d{\bf l}=2\pi L<{\bf \nabla}\alpha>=\frac{\pi L^2}{s}2\pi K$$
where $s$ is the area of the unit cell , $K$ is the number of the unit vortices in the unit cell. That gives 
$|<{\bf \nabla}\alpha>|\approx|K|\pi L/s\gg1/({\sqrt s})$. The factors $\exp({\pm i\alpha})$ strongly oscillate in 
the unit cell due
to the large value of ${\bf \nabla}\alpha$ in the direction parallel to the circumference $r=L$. After the integration in that
direction one get the small factor $\sqrt s/L$ (I assume that the invariant factors have finite derivatives). The integration
in the orthogonal direction gives factor $\approx\sqrt s$ if we take ${\bf k}={\bf k'}$. Thus every unit cell
close to the circumference gives the factor $1/(L s\sqrt s)$ in the transition matrix element. This quantity must be multiplied
by the number of the unit cells along the circumference $\frac{2\pi L}{\sqrt s}$. That gives the total factor $s$ independent on the size $L$. The contribution to the transition matrix element of the inner  circumferences inside the grand circle
is proportional to their number $\frac{L}{\sqrt s}$.

For the second order correction to the ground state energy eq( \ref{delta}) one must take the transition matrix element squared
 and take into account the normalizing factors$1/N^2 $ and the summation over ${\bf k}$ that gives
 $\delta E$ independent on $N$. Thus the second order correction to the ground state energy from  the diagonal terms $E^{0}\sim N$ is vanishingly small and the expression for the thermodynamic 
 energy eq(\ref{effham}) is asimptotically exact at macroscopical large $N$.
 
 The set of the observed densities in the FQHE depends on the value of the gaps, the temperature and the sample purity.
 To take into account these numerous factors is a difficult task and require the intensive numerous calculations to find e.g.
 the unknown periodic function $\beta$ and the type of the lattice ,
 
 It is possible to use Landau argumentation for the explanation of the observed Hall conductance in a weak applied 
 electrical field using a system moving with the electron drift velocity where the electrical field is zero. The electron liquid
 can change it's velocity only  by the birth of the excitations. These excitations are neutral because the consist from
 the electron in the higher empty band and the hole in the filled band. Therefore they have the momentum conservation
 in spite of the magnetic field. The excitation  energy will be $ \epsilon({\bf p})+{\bf pv(E})$. If this quantity is
 positive the excitation's birth is impossible at low enough temperatures and the Hall current is proportional to the
 electron density.
 
 This work is supported by the grant of RFBR, by the program of RAS "Quantum Macrophysics", and the program 
 of Physical Sciences division of RAS " Strongly correlated electron systems".
 Author express his gratitude to V.F. Gantmakher,I.Kolokolov and M.V. Feigelman for the discussion.

\end{document}